\documentclass[aps,pra,reprint,superscriptaddress,floatfix]{revtex4-2}

\usepackage [latin1]{inputenc}
\usepackage[version=4]{mhchem}
\usepackage{graphicx}
\usepackage{amsmath}
\usepackage{hyperref}
\usepackage{siunitx}
\usepackage{bm}
\usepackage[T1]{fontenc} 
\usepackage{dcolumn}
\usepackage{multirow}
\usepackage{xcolor}
\usepackage{array}
\newcommand{\degree}{\ensuremath{^\circ}}%
\usepackage{xr}
\usepackage{microtype}
\usepackage{newtxtext, newtxmath}
\usepackage{xcolor}
\usepackage[pagewise]{lineno}

\let\orgautoref\autoref
\providecommand{\Autoref}{%
  \def\equationautorefname{Equation}%
  \def\figureautorefname{Figure}%
  \def\subfigureautorefname{Figure}%
  \orgautoref}
\let\orgautoref\autoref
\providecommand{\autorefapp}{%
\def\sectionautorefname{App.}%
  \orgautoref}  
\renewcommand{\autoref}{%
  \def\equationautorefname{Eq.}%
  \def\figureautorefname{Fig.}%
  \def\subfigureautorefname{Fig.}%
\def\sectionautorefname{Sec.}%
  \orgautoref}

\begin{document}

\title{Laser-induced Coulomb explosion imaging of \ce{(C6H5Br)2} and \ce{C6H5Br-I2} dimers in helium nanodroplets using the Timepix3}



\author{Constant Schouder}
\affiliation{Department of Chemistry, Aarhus University, 8000 Aarhus C, Denmark}
\author{Adam S.Chatterley}
\affiliation{Department of Chemistry, Aarhus University, 8000 Aarhus C, Denmark}
\author{Melby Johny}
\affiliation{Center for Free-Electron Laser Science, Deutsches Elektronen-Synchrotron DESY, 22607 Hamburg, Germany}
\author{Flora H\"{u}bschmann}
\affiliation{Department of Chemistry, Aarhus University, 8000 Aarhus C, Denmark}
\author{Ahmed F. Al-Refaie}
\affiliation{Center for Free-Electron Laser Science, Deutsches Elektronen-Synchrotron DESY, 22607 Hamburg, Germany}
\author{Florent Calvo}
\affiliation{Universit\'{e} Grenoble Alpes, LIPHY, F-38000 Grenoble, France}
\author{Jochen K\"{u}pper}
\affiliation{Center for Free-Electron Laser Science, Deutsches Elektronen-Synchrotron DESY, 22607 Hamburg, Germany}
\affiliation{Center for Ultrafast Imaging, Universit\"{a}t Hamburg, Luruper Chaussee 149, 22761 Hamburg, Germany}
\affiliation{Department of Physics, Universit\"{a}t Hamburg, Luruper Chaussee 149, 22761 Hamburg, Germany}

\author{Henrik Stapelfeldt}\email{henriks@chem.au.dk}
\affiliation{Department of Chemistry, Aarhus University, 8000 Aarhus C, Denmark}


\date{\today}

\begin{abstract}
We have deduced the structure of the \ce{bromobenzene}--\ce{I2} heterodimer and the \ce{(bromobenzene)2} homodimer inside helium droplets using a combination of laser-induced alignment, Coulomb explosion imaging, and three-dimensional ion imaging. The complexes were fixed in a variety of orientations in the laboratory frame, then in each case multiply ionized by an intense laser pulse. A three dimensional ion imaging detector, including a Timepix3 detector allowed us to measure the correlations between velocity vectors of different fragments and, in conjunction with classical simulations, work backward to the initial structure of the complex prior to explosion. For the heterodimer, we find that the \ce{I2} molecular axis intersects the phenyl ring of the bromobenzene approximately perpendicularly. The homodimer has a stacked parallel structure, with the two bromine atoms pointing in opposite directions. These results illustrate the ability of Coulomb explosion imaging to determine the structure of large complexes, and point the way toward real-time measurements of bimolecular reactions inside helium droplets.

\end{abstract}


\maketitle
\section{Introduction}

Helium nanodroplets offer a unique environment for the creation of weakly-bond complexes~\cite{toennies_superfluid_2004,choi_infrared_2006,yang_helium_2012,doi:10.1080/01442350601087664}. Helium nanodroplets can be, also sequentially, doped with combinations of a broad range of guest molecules\cite{toennies_superfluid_2004,choi_infrared_2006,yang_helium_2012}, and their temperature (0.37 K)~\cite{toennies_superfluid_2004} is low enough to freeze molecules into complexes and clusters that are difficult to achieve with traditional molecular beam techniques. One particularly exciting possibility is the creation of bimolecular complexes in a pre-reactive geometry\cite{doi:10.1080/01442350601087664}, which could be made to react by excitation with a fs laser pulse~\cite{zhong_bimolecular_1996}. To study such a reaction, one needs to first characterize the initial static structure of the complex, and then, ideally, measure the structure as a function of time.

Most studies of molecular complexes in helium droplets have used frequency-resolved spectroscopy to infer structure, in particular IR spectroscopy~\cite{nauta_nonequilibrium_1999,choi_infrared_2006,sulaiman_infrared_2017,verma_infrared_2019}. This is possible because molecules and complexes in droplets tend to show sharp vibrational spectral lines, despite the helium solvent. The inherently limited time resolution of frequency-resolved spectroscopies make them, however, inadequate for measuring how the structure of complexes changes on the natural atomic time-scale. Such studies require structure-sensitive techniques with ps or fs time-resolution.

Recently, we demonstrated that Coulomb explosion imaging (CEI), triggered by intense fs laser pulses, combined with laser-induced alignment and covariance analysis of fragment ion recoil directions, provide an alternative to IR spectroscopy for determining the structure of molecular complexes embedded in He nanodroplets. The method was applied to the homodimers of carbon disulfide~\cite{pickering_alignment_2018}, carbonyl sulfide~\cite{pickering_alignment_2019}, and tetracene~\cite{schouder_structure_2019}. In the gas phase, laser-induced CEI has been a fruitful technique for studying molecular structure~\cite{xu_time-resolved_2016,yatsuhashi_multiple_2018}, including measurements of the handedness of chiral molecules~\cite{pitzer_direct_2013,christensen_using_2015}, determinations of (time-dependent) internuclear wave functions of diatomic molecules and atomic or molecular dimers~\cite{stapelfeldt_wave_1995,petersen_control_2004,ergler_spatiotemporal_2006,zeller_imaging_2016,schouder_pra_2020}, and imaging of intramolecular motions and isomerization dynamics in real-time~\cite{madsen_manipulating_2009,ibrahim_tabletop_2014,christensen_dynamic_2014}.

Conceptually, CEI is simple: an intense laser pulse multiply ionizes molecules, which then break up into cationic fragments. Applying the axial recoil approximation, the original molecular structure can then be reconstructed from the velocity vectors of the fragments. Coincidence and covariance techniques are essential here, as they allow the velocity vectors from explosion of a single system to be related to each other~\cite{jagutzki_multiple_2002,frasinski_covariance_2016,gagnon_coincidence_2008,hansen_control_2012}. For the structural determination of carbon disulfide, carbonyl sulfide, and tetracene dimers, we also used laser-induced alignment to fix the dimers in the laboratory frame~\cite{pickering_alignment_2019,schouder_structure_2019}
.
Laser-induced alignment is the practice of using moderately intense non-resonant laser pulses to fix a molecule's spatial orientation in the laboratory frame through the polarizability interaction~\cite{stapelfeldt_colloquium:_2003,fleischer_molecular_2012,kumarappan_aligning_2007}. Alignment is key to CEI structural determination, because it optimizes the information content about the molecular structure extractable from the fragment velocity vectors. Also, it allows us to relate the velocity vectors of different fragments to the most, second most, or least polarizable axes of the complex. For instance, for the tetracene dimer we were able to tell how the two monomers were oriented with respect to the polarizability axes (which themselves are dependent on the dimer structure), and from this deduced a parallel stacked structure~\cite{schouder_structure_2019}.



The limitation of the technique introduced in \cite{schouder_structure_2019} was that it was only applicable to homodimers, because it depends on gating the imaging detector in time to detect only ion species with similar mass-to-charge, $m/z$, values, the \ce{tetracene+} ion in that case. However, if one uses a three-dimensional imaging detector, which records the time as well as the position of each ion hit, it is possible to correlate events between ion species with different $m/z$ ratios. Such detectors have traditionally used either delay lines~\cite{jagutzki_multiple_2002,dorner_cold_2000,ullrich_recoil-ion_2003} or correlated photon counters~\cite{lee_coincidence_2014,weeraratna_demonstration_2018} to determine ion timings, with the drawback that these detectors are limited to a few ion events per laser shot, so data collection is rather laborious. Recently, a new generation of 3D detectors has emerged in the form of 'cameras' whose pixels record a time-stamp of each ion event. Ion imaging experiments employing either the PImMS~\cite{nomerotski_pixel_2010,nomerotski_pixel_2011,john_pimms_2012} or Timepix~\cite{poikela_timepix3_2014,zhao_coincidence_2017} devices are able to measure the time-stamp and location of hundreds of events in every laser shot, significantly enhancing data collection times and enabling statistical analysis of many fragment correlations.

Here, we have used a Timepix3 camera to overcome the homodimer limitation, and measure the structure of the bromobenzene -- iodine (\ce{BrPh}--\ce{I2}) heterodimer inside a helium droplet. We do this by measuring correlations between the emission directions of \ce{Br+} and \ce{I+} fragments following Coulomb explosion, for complexes fixed in space using laser-induced alignment. We find a structure where the \ce{I2} molecule is nearly perpendicular to the BrPh aromatic ring very similar to related gas phase complexes \cite{walker_structure_1995,grozema_iodinebenzene_1999,kiviniemi_iodinebenzene_2009}. Additionally, we have determined the structure of the \ce{(BrPh)2} homodimer 'for free', as a number of these complexes are also present in our helium droplets, and the detection system simultaneously records their recoil data while the heterodimer CEI is being recorded. We find a parallel displaced structure, with the two Br atoms pointing away from each other. These experiments demonstrate the power of combining alignment, 3D ion imaging, and CEI to simultaneously measure the structures of multiple complexes, and point the way toward measurement of real-time structural changes of loosely bound complexes.

\section{Experimental setup}
The general experimental setup has been described in detail before\cite{shepperson_strongly_2017,schouder_structure_2019}; the two key differences here are the use of two separate doping cells for creation of heterodimers in He droplets, and the replacement of the CCD camera by a Timepix3 detector. The helium nanodroplets are produced by expanding 25 bar He into vacuum through a \SI{5}{\micro m} nozzle, cooled to 14 K. This produces liquid helium droplets with a mean size of 5000 He atoms \cite{toennies_superfluid_2004}. Molecular dimers are formed by sending the droplets through two doping cells, the first containing BrPh vapor and the second \ce{I2} vapor. The vapor pressure inside the two doping cells is controlled by needle valves connected to external room temperature reservoirs, and is set to optimize signal from \ce{BrPh-I2} heterodimers, while minimizing signal from unsolvated gas phase molecules. In practice, this amounts to turning both vapor pressures as low as possible, while still observing dimer signals.

The doped helium droplets enter the target region, where they are intersected at
\SI{90}{\degree} by two pulsed laser beams. The pulses in the first beam are used to adiabatically align the dimers. Each pulse turned on in $\sim$\SI{120}{ps} and off in $\sim$\SI{10}{ps}. The asymmetric shape is obtained by  spectral truncation of the uncompressed pulses from a regenerative amplified fs laser system~\cite{chatterley_communication:_2018,mullins2020picosecond}. The alignment pulse parameters are: $\lambda_\textrm{align} = \SI{800}{nm}$, $\omega_0=\SI{40}{\mu m}$, and peak intensity I$_\textrm{align}\sim\SI{3e11}{W/cm^2}$. The second laser beam contains the probe pulses used to multiply ionize the molecules in the droplets and thereby trigger Coulomb explosion of the dimers. The probe pulse parameters are: $\lambda_\textrm{probe}=\SI{400}{nm}$, $\tau_\textrm{probe}= \SI{40}{fs}$ (FWHM), $\omega_0=\SI{25}{\micro m}$, and I$_\textrm{probe}\sim\SI{2e14}{W/cm^2}$. Each probe pulse is sent  $\sim$\SI{5}{ps} after the truncation of the alignment pulse. This timing is chosen such that the dimers are both well-aligned and that the alignment field is negligible at the time of probing~\cite{chatterley_long-lasting_2019}. The spot size of the probe beam is considerably smaller than the spot size of the alignment beam to minimize focal volume effects. The polarization of the probe laser pulse is linear and orthogonal to the detector plane, while the ellipticity and plane of polarization of the alignment pulse is varied to control how the dimers are aligned. In all cases, the effect of the alignment pulse is to coerce the most polarizable axis (MPA) of the dimers to coincide with its major polarization axis. If the alignment pulse is linearly polarized, then the complex rotates freely around the polarization axis. When elliptically polarized laser pulses are used, the second most polarizable axis of the dimers is constrained to the minor polarization axis, and the complex is 3D aligned~\cite{larsen_three_2000,chatterley_three-dimensional_2017}.

Following fragmentation, the velocities of the resulting ions are projected onto a 2D detector by a velocity map imaging spectrometer. Usually, such apparatus is temporally gated to only accept ions with one specific $m/z$ value at the time. With the TimePix3 detector~\cite{poikela_timepix3_2014} and its SPIDR read out~\cite{Visser_2015}, both the spatial coordinates of each ion impact event, and its time-of-flight with a temporal resolution of a few ns are measured. Each ion impact results in a small cluster of pixels registering events containing the pixel coordinates and the time of the hits. This data is recorded and clusters of pixels are separated and centroided in time and space using the Pymepix software to give a list of ion hits~\cite{al-refaie_pymepix_2019}. The time of each hit is used to differentiate fragments with different $m/z$, values, while the spatial coordinates give the projected velocities of the ionic fragments.

The detector is gated to ignore all events with $m/z < 63$ u, firstly to avoid damage to the detector from the very significant number of light ions, including \ce{He+} ions and residual water in the chamber, and secondly to ensure that the centroiding system could keep up with the incoming data rate for online data analysis.

The outcome of these experiments is a set of 3D data with two polar coordinates $r$ and $\theta$ (transformed from the Cartesian pixels), and one time coordinate, $t$. For each ion event, $r$ and $\theta$ give the projection of the velocity vector, $v$, onto the 2D detector, and $t$ gives the time-of-flight and thereby the $m/z$ value. Throughout the manuscript, $r$ will be given in relative units of pixels, as read-off from the camera, and $\theta$ in degrees with respect to a space-fixed axis (the laser propagation axis). The ion species detected will be denoted by a lower index, e.g., $r_{\rm Br^+}$.

Our primary analytical tool is covariance analysis, which gives the likelihood of two events being correlated as a function of one of their coordinates ($r, \theta$ or $t$)~\cite{frasinski_covariance_2016}. Formally, the covariance is defined as $C_{UV} = \langle UV \rangle - \langle U\rangle\langle V\rangle$, where $U$ and $V$ are the histograms of the coordinates for events of the two ion species being considered, and angle brackets denote expectation values. For the mass spectrum covariance, \autoref{sec:TOF}, $U=V$ and is the transformed time-of-flight of every ion event. For the angular and radial covariance of a single species, $U$ and $V$ are each histograms of the respective coordinate, but only for events whose $m/z$ value is equal to the ion species of interest.

\section{Results and analysis}

\subsection{Mass spectra}
\label{sec:TOF}

\begin{figure}[bt]
\centerline{\includegraphics[width = \columnwidth]{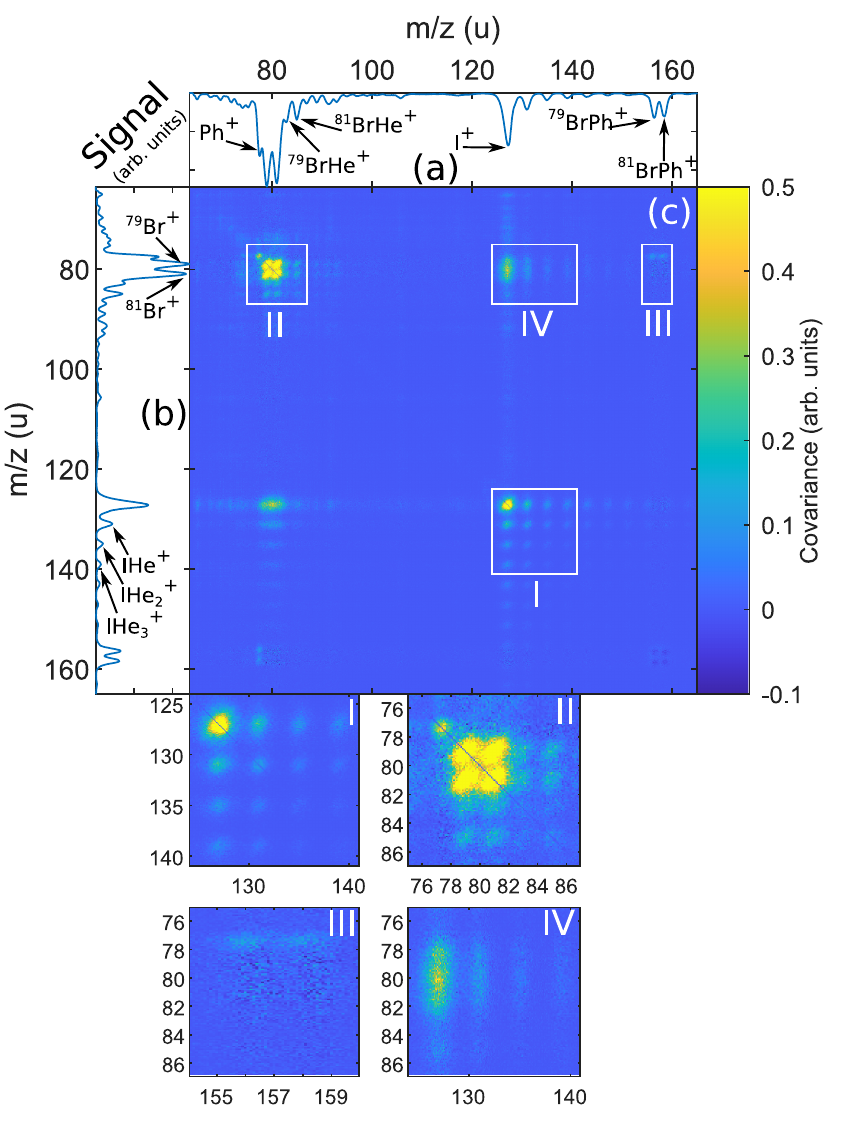}}
\caption{The mass spectrum and mass spectrum covariance map of helium droplets doped with both \ce{BrPh} and \ce{I2} molecules, following Coulomb explosion. (a) and (b) show the 1D mass spectrum, while (c) shows the covariance of these spectra. Positive peaks in the covariance indicate fragments that arrive alongside each other. Four regions of interest, I, II, III, and IV, are highlighted and expanded, see text for details.}
\label{MASS}
\end{figure}

The time axis of our 3D dataset is the mass spectrum after laser ionization when both doping cells are opened, and the covariance map of these data shows which fragments appear in conjunction with each other~\cite{frasinski_covariance_1989}. We shall focus only on the mass region of 70--160 u (\autoref{MASS}), which spans \ce{Ph+}, \ce{Br+} and \ce{BrPh+} fragment ions from BrPh molecules, and \ce{I+} ions from \ce{I2} molecules, as these are the most useful for structural determination. The full mass spectrum is shown in \autoref{MASSTOT}. Although alignment has little effect on the yield of the mass spectra, we note for completeness that the data in \autoref{MASS} are recorded with the alignment pulse linearly polarized parallel to the detector plane.

Laser-induced ionization results in a variety of fragment ions, but for structural determination purposes we are only interested in fragments that appear in conjunction with another fragment. These appear as positive peaks in the mass spectrum covariance map in \autoref{MASS}. Peaks along the diagonal correspond to fragments correlated with an identical partner, while off-diagonal peaks reveal fragments correlated with a different partner. Four groups of peaks are visible in the covariance map, labeled I--IV in \autoref{MASS}. Note that the covariance map is symmetric about the diagonal. Region I is a cluster of peaks originating at (127,127) with weaker peaks shifted periodically by 4 u either horizontally or vertically. These peaks correspond to Coulomb explosion of an \ce{I2} molecule in a single He droplet into two \ce{I+} ions each of which may pick up one or more He atoms as it escapes the droplet~\cite{christiansen_laser-induced_2016}. This effect of helium 'snowball' formation is consistently observed when \ce{I+} ions are produced from laser-induced Coulomb explosion or photodissociation of molecules inside helium droplets~\cite{braun_photodissociation_2007-1,christiansen_laser-induced_2016}.

Region II is also a cluster of evenly spaced peaks extending beyond the diagonal. The main peaks at (79,79), (81,79) and (81,81), correspond to correlated pairs of \ce{^{79}Br+} and \ce{^{81}Br+} ions (the natural abundances of \ce{^{79}Br} and \ce{^{81}Br} are almost the same) produced by Coulomb explosion of \ce{BrPh} molecules, with the possibility for He atom pickup in the same manner as for the \ce{I+} ions. An additional peak at (77,77) corresponds to a correlated pair of \ce{Ph+} ions. As each \ce{BrPh} molecule contains only one Br atom and one Ph fragment, the covariance peaks in region II must stem from He droplets doped with a \ce{(BrPh)_n} oligomer with n $\geq 2$, created when a droplet picks up multiple \ce{BrPh} molecules. Further confirmation of the formation of \ce{(BrPh)_n} comes from the two peaks in region III, which correspond to a \ce{Ph+} ion detected in coincidence with a \ce{^{79}BrPh+} or \ce{^{81}BrPh+} parent ion. We consider that the major contribution of the correlation signal originated from \ce{(BrPh)2} as the doping cell pressure was kept as low as possible to minimize the contribution of larger oligomers.

The peaks in region IV show \ce{Br+} ions arriving in conjunction with \ce{I+} ions. Such a correlation must originate from ionization of a complex containing at least one \ce{I2} and one \ce{BrPh} molecule, the simplest of which is the \ce{BrPh-I2} heterodimer. The mass spectrum covariance map leads us to conclude that we have produced both \ce{(BrPh)2} homodimers and \ce{BrPh-I2} heterodimers. In principle, parts of the correlation signals could also come from larger oligomers, but the doping cell pressure was kept as low as possible to minimize this contribution.

\subsection{Heterodimer}
\label{sec:heterodimer}
\begin{figure}[bt]
\centerline{\includegraphics[width = \columnwidth]{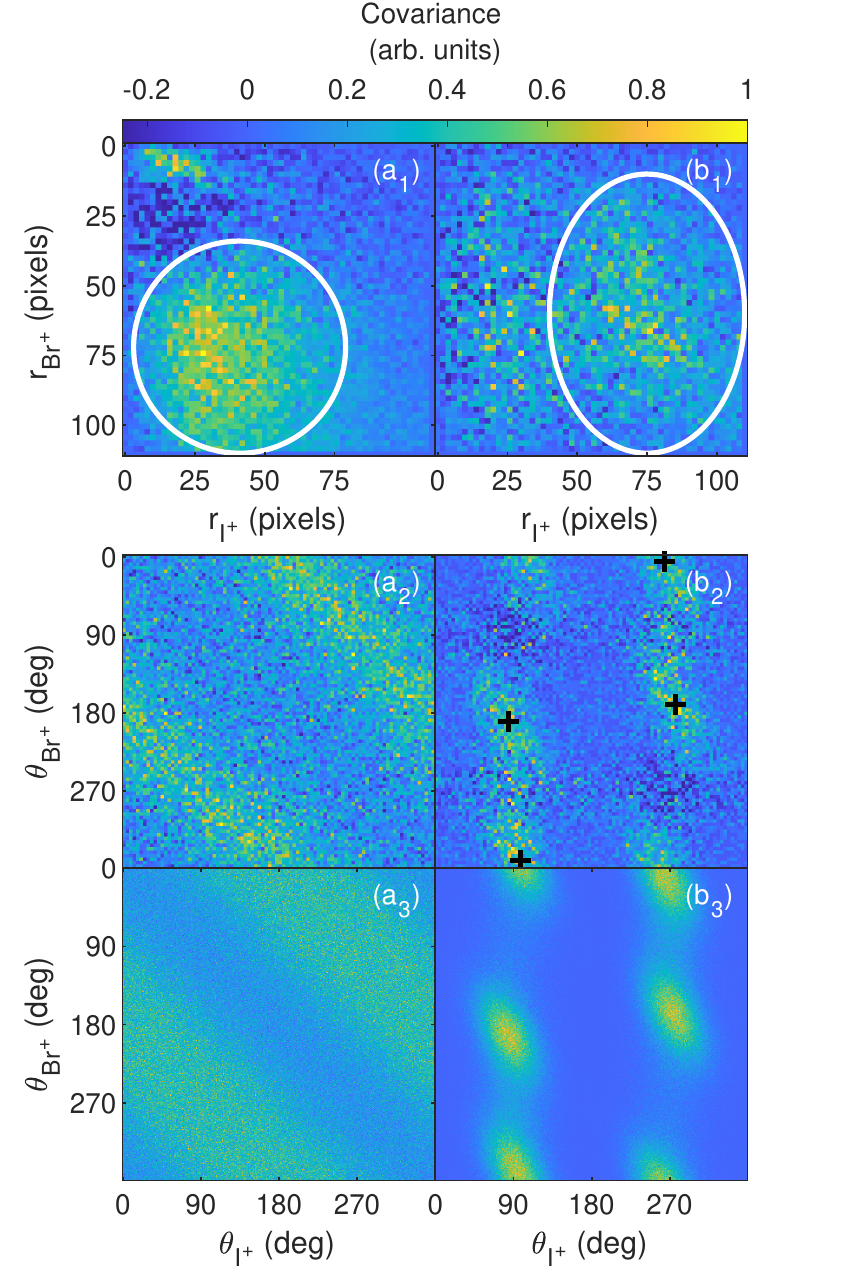}}
\caption{(a$_1$)-(b$_1$): Radial covariance map between \ce{I+} and \ce{Br+} ions. (a$_2$)-(b$_2$): Angular covariance map for the \ce{I+} and \ce{Br+} ions detected in the ovals shown in (a$_1$) and (b$_1$), the black crosses mark the position of the islands' center listed in \autoref{centreTAB}. (a$_3$)-(b$_3$): Simulated angular covariance map (see text). In the left (right) column the MPA is aligned perpendicular (parallel) to the detector plane.}
\label{I2BRPHCOV}
\end{figure}

The mass spectrum covariance map shows that we form \ce{BrPh-I2} heterodimers, but does not give much information about their structure. For this, we turn to the velocity vector covariance between \ce{I+} and \ce{Br+} fragments, in particular the radial and the angular covariance maps~\cite{christiansen_laser-induced_2016}. The radial covariance map allows us to identify correlation regions that are related to the ionization of the heterodimer. The angular covariance map is then computed for ions in these correlation regions and analyzed to identify the recoil angle of the correlated fragments under study. We choose to study the correlations between the \ce{I+} and \ce{Br+} fragment ions because they offer the most direct insight into the complex structure: if we assume axial recoil then the \ce{I+} velocity reveals the I--I bond axis, while the \ce{Br+} velocity reveals the orientation of the C--Br axis of the \ce{BrPh} molecule. For these measurements, it is essential to fix the alignment of the dimer in the laboratory frame for two reasons. Firstly, alignment restricts the range of laboratory recoil angles of the fragments, which sharpens the peaks in the covariance maps and simplifies analysis. Second, the direction in which the dimer aligns is determined by its polarizability tensor, which in turn is determined by the dimer structure. Thus, observing how the dimer aligns for a given polarization state of the alignment field reveals information about the structure of the complex~\cite{pickering_alignment_2019,schouder_structure_2019}. For the measurements presented in \autoref{I2BRPHCOV}, we use one-dimensional alignment with the major polarizability axis (MPA) aligned either perpendicular (left column) or parallel (right column) to the detector plane. \Autoref{I2BRPHCOV}(a$_1$)-(b$_1$) shows the radial covariance map and \autoref{I2BRPHCOV}(a$_2$)-(b$_2$) the angular covariance map of (\ce{I+}, \ce{Br+}) ion pairs obtained from the ion images recorded. Furthermore, to reduce the contribution from oligomers larger than the heterodimer, we discard the ion images from those laser shots where more than one \ce{Br+} ion is detected. For example, in the case of an alignment laser parallel to the detector plane (see \autoref{I2BRPHCOV} panels (b$_1$) and (b$_2$)), there are 535757 lasershots that lead to the detection of at least one \ce{Br+} ion, out of which 393088 contained only one.

For each of the \ce{I+} -- \ce{Br+} covariance maps we shall now consider what insights they give us about the structure of the complex.
The radial covariance map with the MPA aligned perpendicular to the detector, displayed in \autoref{I2BRPHCOV}(a$_1$), shows that high-velocity \ce{Br+} ions, detected at radii between 40 and 110 pixels, are correlated with lower velocity \ce{I+} ions detected at radii between 10-50 pixels. This indicates that the velocity vector of the \ce{Br+} ions, and thus the C--Br axis of the parent \ce{BrPh} molecule, is in or close to the detector plane while the velocity vector of the \ce{I+} ions, and thus the I--I axis, is close to perpendicular to the detector plane. The radial covariance map thus points to a dimer structure where the MPA is along the I--I axis and perpendicular to the C--Br axis. We also determined the angular covariance map for the events with radial covariance contained in the white oval in \autoref{I2BRPHCOV}(a$_1$). \Autoref{I2BRPHCOV}(a$_2$) shows two broad stripes of positive covariance signal, showing correlation for (\ce{I+},\ce{Br+}) ion pairs where $\theta_{\rm Br^+} = \theta_{\rm I^+} + 180\degree$, i.e. the two fragments recoil back-to-back. This is consistent with the dimer structure tentatively suggested from the radial covariance map but does not add any further insight.


More detailed information comes from the radial and angular covariance maps when the dimer is aligned with the MPA, which we now believe to be close to the I--I axis, parallel to the detector plane. \Autoref{I2BRPHCOV} (b$_1$) shows the radial covariance map for this alignment. The most prominent feature is a positive region centered around $(r_{\rm Br^+} = 60, r_{\rm I^+} = 75)$ pixels. When compared to \autoref{I2BRPHCOV}(a$_1$), it shows a decrease and an increase of the kinetic energy release in the detector plane for the \ce{Br+} and \ce{I+} fragments respectively. Hence similar arguments apply and we conclude that for this alignment geometry $v_{\rm I^+}$ is more in the plane of the detector than $v_{\rm Br^+}$, which is consistent with the assessment that the I--I axis coincides with the MPA. However, the angle between the I--I and C--Br axes is still unclear. The angular covariance in this alignment geometry \autoref{I2BRPHCOV}(b$_2$), contains most of the information on this parameter. Here, we see four separate islands (note that the axes wrap around), at locations close to (\SI{90}{\degree}, \SI{180}{\degree}), (\SI{90}{\degree}, \SI{0}{\degree}), (\SI{270}{\degree}, \SI{180}{\degree}) and (\SI{270}{\degree}, \SI{0}{\degree}), although not centered exactly on these spots.  The positions of the islands indicate that the \ce{I+} fragment is flying in a direction close to the alignment laser polarization (\SI{90}{\degree} or, equivalently, \SI{270}{\degree}), while the \ce{Br+} fragment flies nearly perpendicular to it, confirming the previous observations. Putting all the covariance observations together, we arrive at a qualitative structure with the I--I axis perpendicular to the C--Br axis.
In panel (b$_1$) of \Autoref{I2BRPHCOV}, a fainter contribution appears for lower radius $(r_{\rm Br^+} = 60, r_{\rm I^+} = 25)$ which is interpreted as originating from the \ce{I+} ion pointing toward the phenyl ring. The angular covariance map of this region is featureless and does not show a confined distribution. which can be expected from the large scattering that this fragment can undergo compared to the outer one.

\begin{table}[bt]
\centering
\begin{tabular}{|c|c|}
\hline
Island & ($\mu_{\theta_{\textrm{I}^+}} \pm \sigma_{\theta_{\textrm{I}^+}}$ , $\mu_{\theta_{\textrm{Br}^+}} \pm \sigma_{\theta_{\textrm{Br}^+}} $)\\
\hline
(1)&($    96.2\pm 1.1 $ , $  349.2\pm 3.2 $)\\
(2)&($  262.8\pm  1.3 $ , $    5.8\pm 3.1 $)\\
(3)&($   82.7\pm 1.5 $ , $  189.9\pm 4.6$)\\
(4)&($  275.0\pm 0.9 $ , $  170.0\pm 3.8 $)\\
\hline
\end{tabular}
\caption{The center positions of the four islands in \autoref{I2BRPHCOV}(b$_2$), determined using a clustering algorithm detailed in \autorefapp{sec:PeakCenter}.
\label{centreTAB}}
\end{table}
Quantitative details of the structure come from comparison of the angular covariance with simulations. The first step is to precisely identify the center of each island in \autoref{I2BRPHCOV}(b$_2$), ($\mu_{\theta_{\rm I^+}}$ and $\mu_{\theta_{\rm Br^+}}$), using a clustering algorithm detailed in the \autorefapp{sec:PeakCenter} and shown in \autoref{CENTERS}. The centers, and their uncertainties, are shown in \autoref{centreTAB}. Next, we construct trial structures, simulate their angular covariance matrices, and then extract the center position of the islands as in the experimental data. We then use an optimization algorithm to minimize the difference between the island centers of the experimental and simulated covariance maps by varying the trial structure. For each trial structure we first assign a charge distribution to each atom (see below), then classically simulate the atomic motion, assuming only Coulombic forces are significant. The resultant velocity vectors are then transformed into an ensemble distributed around the alignment polarization vector, taking into account the imperfect alignment of the experiment. To account for scattering of the fragment ions as they move out of the helium droplet~\cite{christensen_deconvoluting_2016,shepperson_strongly_2017} we convolute the final velocities of the \ce{Br+} and \ce{I+} ions with a Gaussian function. The angular covariance maps are then computed, and compared to the experimental results. The structure of the dimer is then varied until the simulated angular covariance matches the experiment.

\begin{figure}[bt]
\centerline{\includegraphics[width = \columnwidth]{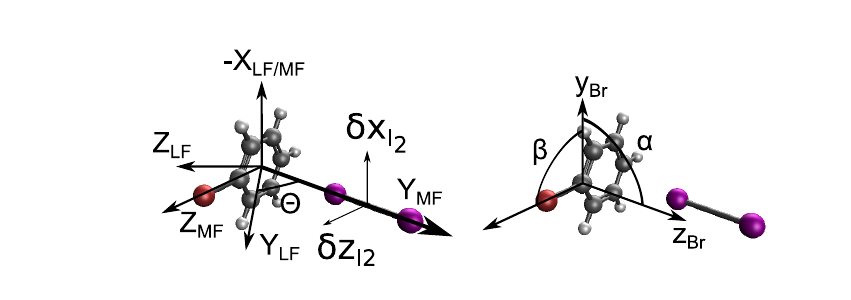}}
\caption{The starting structure for the optimization of the \ce{BrPh-I2} complex, along with the structural parameters that were optimized.
\label{COORDS}}
\end{figure}

There are 6 parameters to describe the structure for optimization, (highlighted in \autoref{COORDS}: (1) An angle $\theta$ gives the the angle between the \ce{I2} axis and the lab-frame major polarization axis of the alignment laser ($Y_{LF}$); and rotates the whole dimer around the $Z_{LF}$ axis of the laboratory frame; (2-3) the \ce{I2} molecule is allowed to move in the $X-Z$ plane of the molecular frame with parameters $\delta x_{I_2}$ and $\delta z_{I_2}$ while the distance between the benzene ring and the closest iodine atom is initially set to \SI{3.5}{\angstrom}; (4-6) a set of three Proper Euler angles ($\alpha, \beta, \gamma$) associated respectively with the sequence of rotation axes $\left(x_{Br}-z_{Br}-x_{Br}\right)$ permits free rotation of the bromobenzene molecule in the laboratory frame. Each monomer of the complex is kept rigid in its equilibrium geometry (computed with the $\omega$B97X-D method~\cite{B810189B} with aug-cc-pVTZ basis~\cite{doi:10.1063/1.456153} set using Gaussian09~\cite{g09}). The starting geometry for the optimization is the experimentally deduced qualitative structure, with the \ce{I2} molecule perpendicular to the phenyl ring. The exact starting geometry is given in \autoref{Table:InputGeometry} in the appendix.


The simulations also depend on the distribution of charges within the complex, following Coulomb explosion. Unfortunately, we do not know precisely how the ionization and fragmentation occur. However, as we only focus experimentally on the \ce{I+} and \ce{Br+} fragments, we can select specific ionization and fragmentation cases that must fulfill: (i) the double ionization of the \ce{I2} molecule fragmenting into two single charged \ce{I+} ions. (ii) the fragmentation of BrPh leading to a single \ce{Br+} ion, along with some other ionic fragments.
We consider the following feasible fragmentation pathways for the BrPh molecule, these channels are chosen such that the mass and the charges of the ionic fragments are sufficiently different to lead to a broader distribution of the plausible Coulomb explosion dynamics:

\begin{align}
\tag{a}\ce{C6H5Br^{4+} &\rightarrow Br^{+} + 2C2H2^{+} + CH^{+} + C}\\
\tag{b}\ce{C6H5Br^{6+} &\rightarrow Br^{+} + 2C2H2^{2+} + CH^{+} + C}\\
\tag{c}\ce{C6H5Br^{6+} &\rightarrow Br^{+} + 5CH^{+}  + C }\\
\tag{d}\ce{C6H5Br^{2+} &\rightarrow Br^{+} +C6H5^{+}}\\
\tag{e}\ce{C6H5Br^{3+} &\rightarrow Br^{+} +C6H5^{2+}}
\end{align}

For each of the molecular fragments from these five pathways we assign the charge on each atom by fitting an electrostatic potential using the fluctuating charge method ($\omega$B97X-D method~\cite{B810189B} with aug-pcseg-n basis set~\cite{doi:10.1021/ct401026a}). For each fragmentation pathway, we now have a charge assigned to each atom, which are listed in \autoref{Table:Conf} in the appendix. The use of partial charges and Coulombic force to simulate the dynamics will likely underestimate/overestimate the interaction between each ionic fragment as the non-Coulombic part of the interaction will not be taken into account\cite{PhysRevA.102.063125}. However, its directionality should not be too affected by the inclusion of these effects thanks to the use of partial charges to reproduce the electrostatic potential.

\begin{figure}[bt]
\centerline{\includegraphics[scale = 1]{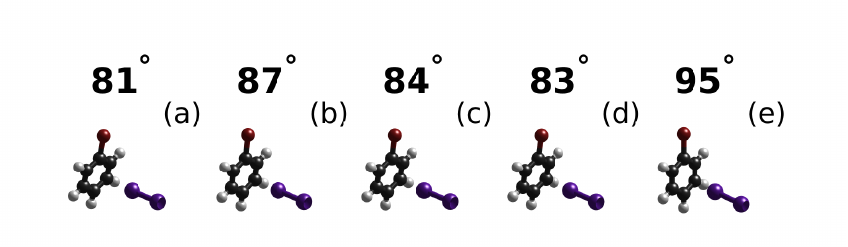}}
\caption{Structures fulfilling experimental velocity recoil for several charge and fragmentation schemes,label is associated to a particular fragmentation pattern and a specific charge distribution detailed in the text. The full parameters are given in \autoref{Table:Conf} in the appendix. The angle on top of each panel is the relative angle between the \ce{I2} axis and the \ce{C-Br} axis. 
 }
\label{HETEROSTRUCTURES}
\end{figure}

We find that the optimization converges well on the experimental angular covariance for all five fragmentation pathways. Figure \ref{I2BRPHCOV} (a$_3$) and (b$_3$) show the simulated angular covariance for the case of fragmentation pathway (a), although the covariance plots simulated for the other four fragmentation channels look essentially the same. The agreement is excellent, although we note that the shape of the islands is controlled mostly by the assumptions of imperfect alignment and non-axial recoil. However, the island centers, which are the target for the optimization, depend solely on the initial structure and fragmentation pathway. The geometries that the optimization finds do not deviate far from our initial input, with the main change being that the angle between the I--I and C--Br axes now varies from 70\degree to 95\degree, depending on the fragmentation pathway choice.  The five resultant structures, and this angle, are shown in \autoref{HETEROSTRUCTURES}. 
These structures are local minimum in the optimization landscape and should not be considered as trustworthy predictions of the real geometry. The optimization algorithm can lead to an almost perfect replica of the experimental data over a broad distribution of initial geometry limiting the accuracy of the prediction and the interpretation of the final given structures found. In order to improve the modeling and to reduce the optimization landscape, more observables would be needed such as \ce{C+} and \ce{H+}. Nevertheless, it provides a starting guess for further inquiries. 

To test these classical predictions, we calculate the polarizability tensor of each structure, shown in \autoref{table:Polarizability} in the appendix. We use the $\omega$B97X-D method~\cite{B810189B} with Def2QZVPP~\cite{B508541A} as a basis set. The largest component of the polarizability tensor identifies the axis of the complex that aligns along the alignment laser polarization axis. It is then possible to extract the relative angles between the I--I and C--Br axes using the alignment laser polarization as a reference. These angles can then be compared with the prediction from classical simulations. The results are displayed in \autoref{table:AngleClassicalQuantum} in the appendix and allow us to discard structures (a) and (e), because they show a large offset in the predicted angles between the classical calculations and the quantum approach. The most likely explanation is that the classical calculations are not able to predict correctly the initial structure from these two charge/fragment distributions.
As only structures (b), (c) and (d) show a consistent behavior between classical and quantum mechanics, we believe it provides the best estimate for the angle. Nevertheless, the large spread used to reproduce the island shape makes the estimation of its spread difficult. Therefore, we use the upper value of the experimental noise used in the simulations, $20^\circ$, and give an estimate of $85 \pm 20\degree$ for the angles between the I--I and C--Br axes.

Regarding the position of the \ce{I2} molecule relative to the molecular plane of \ce{BrPh}, we find that the agreement between the simulations and the experiment is roughly equal whether the \ce{I2} is located above the center of the phenyl ring, above a carbon atom, or above a C--C bond. As such, the current experimental observables only allows us to estimate the angle between the \ce{I2} molecule and the C--Br axis, but not the spatial location of the \ce{I2} molecule above the \ce{BrPh} molecular plane. Insight into the latter are obtained by theoretical calculations of the stable structures of the heterodimer in the gas phase, see \autorefapp{sec:theory} for details. Four stable structures are identified as shown in \autoref{Figure:StructureTheory}. For the two most stables structures, the angle between the I--I and C--Br axes is $81.8^\circ$ and $93.0^\circ$ and the I--I axis lies above one carbon atom. The two others structures are higher in energy by about 70 meV and the I--I axis is parallel to the molecular plane of \ce{BrPh}. Therefore, the experimentally found result is in good agreement with the two most stable structures predicted by the gas phase geometry optimization.

\subsection{Homodimer}
\begin{figure}[bt]
\centerline{\includegraphics[width = \columnwidth]{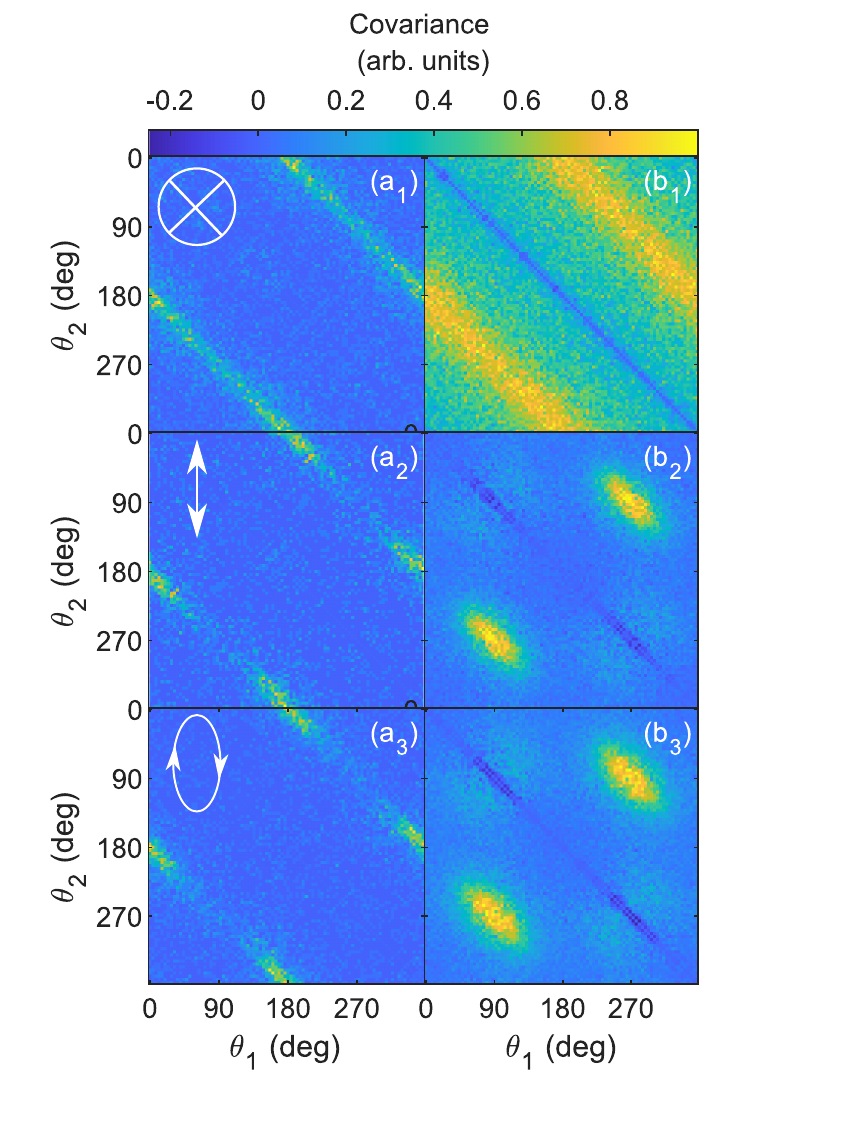}}
\caption{Angular covariance maps of \ce{BrPh+} parent ions (left column) and \ce{Br+} ions (right column), after Coulomb explosion of aligned \ce{(BrPh)2} dimers. The alignment geometry is indicated on the left of each row. In all cases, only ions with significant kinetic energy were considered.
\label{BRCOV}}
\end{figure}

In addition to the \ce{BrPh-I2} heterodimer, we have also determined the structure of the \ce{(BrPh)2} homodimer, using a similar methodology. Most of the structural information comes from the angular covariance maps of either \ce{BrPh+} or \ce{Br+} fragments with themselves. First, the parent ion covariance maps are consistent with four different dimer configurations. The parent ion covariance is calculated by only considering the energetic ions originating from Coulomb explosion of the dimer, similar to the approach used for the \ce{CS2}~\cite{pickering_alignment_2019} and the tetracene dimer~\cite{schouder_structure_2019}. In practice, we implement this by selecting parent ions with a radius larger than 11 pixels. The \ce{Br+} fragments allow us to directly observe the relative orientation of the two monomers, and hence deduce the final structure.

\begin{figure}[bt]
\centerline{\includegraphics[width = \columnwidth]{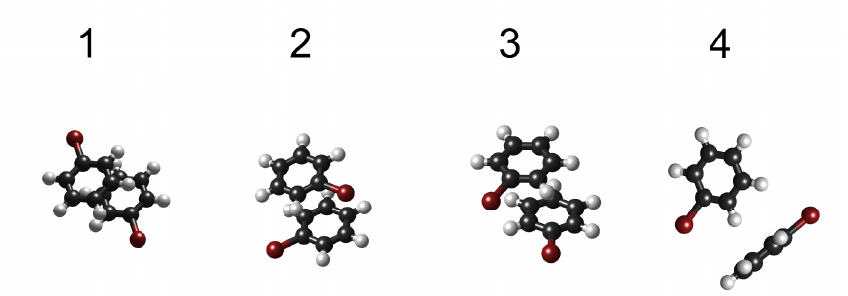}}
\caption{Conformations of the \ce{BrPh2} consistent with the experimental angular covariance maps from \autoref{BRCOV}. The angle difference between the directions of the C--Br axes is 1: 180$^\circ$, 2: 120$^\circ$, 3: 30$^\circ$ and 4: 180$^\circ$.}
\label{BRCONF}
\end{figure}

The \ce{BrPh+} -- \ce{BrPh+} angular covariance map is given in \autoref{BRCOV} (a1)--(a3), for the three different alignment polarization geometries. In all cases, the two fragments recoil with a $180^\circ$ angle between them, as expected for a two-body breakup.  However, different alignment geometries result in different localizations of the angles, which tells us where the MPA is relative to the two monomers. When the MPA is aligned perpendicular to the detector [\autoref{BRCOV} (a1)] shows an isotropic distribution. However, alignment of the MPA parallel to the detector [\autoref{BRCOV} (a2)] shows confinement to two islands centered at (0\degree, 180\degree) and equivalently (180\degree, 360\degree). Alignment with an elliptically polarized laser pulse, which also confines the second most polarizable axis perpendicular to the detector, shows an angular covariance map [\autoref{BRCOV} (a3)] almost identical to the one in \autoref{BRCOV} (a2). In analogy with the analysis in recent studies on the dimer of tetracene~\cite{schouder_structure_2019}, also a planar molecule, we identify four possible dimer structures that are consistent with the observed angular covariance maps. Three of them have a stacked parallel displaced geometry (1,2,3), and one has a T-shaped geometry (4), as shown in \autoref{BRCONF}.

We can distinguish between these four plausible structures with the help of the \ce{Br+} -- \ce{Br+} angular covariance, shown in \autoref{BRCOV}(b1)--(b3) for the three different alignment polarizations. In all cases, the two \ce{Br+} ions recoil in opposite directions with a $180^\circ$ relative angle. In the two cases where the MPA is parallel to the detector [\autoref{BRCOV} (b2)--(b3)], the \ce{Br+} ions are localized at $90^\circ$ and 270\degree, i.e. along the polarization direction of the alignment pulse. This means that the MPA in the dimer is parallel to the C -- Br axes. Inspecting the four structures in \autoref{BRCONF}, the immediate impression is that this is only the case for structure 1. Simulating the recoil direction from Coulomb explosion of the four different structures corroborates that only conformation 1 leads to recoiling \ce{Br+} ions that reproduce the observed experimental covariance maps as shown in \autoref{BrCov}. Again, we compare our experimental findings with predictions from quantum chemistry as discussed in \autorefapp{sec:theory}. Three displaced parallel structures are found, the lowest one shows a parallel displaced structure with the bromine opposing to one another while the other two, also parallel displaced, show an angle of 120$^\circ$ and 30$^\circ$ between the two C--Br axes and have an energy respectively of 15 and 39 meV above in energy. Therefore, only the first one seems to be consistent with our experimental data.

\section{Discussion and Outlook}
\subsection{Structure}

The structure that we arrive at for the \ce{BrPh-I2} complex inside a helium droplet is one where the \ce{I2} molecule is nearly perpendicular in the phenyl ring. To our knowledge, there has been no previous investigations into this complex. However, the related \ce{I2 - C6H6} complex has been rather well studied, because it serves as a model system for charge transfer processes\cite{walker_structure_1995,grozema_iodinebenzene_1999,kiviniemi_iodinebenzene_2009,deboer_molecular_1996,deboer_photodissociation_1996,weng_resonance_2006,cheng_femtosecond_1996}. Matrix isolation spectroscopy found that the \ce{I2} molecule is exactly perpendicular to the benzene plane, and lies directly above one of the C--C bonds \cite{kiviniemi_iodinebenzene_2009}, in agreement with previous theoretical calculations \cite{grozema_iodinebenzene_1999}. From our measurements of \ce{BrPh-I2} in He droplets, it appears that the angle between the two molecules is slightly less than perpendicular, at around 85\degree\cite{doi:10.1063/1.472710}. This is perhaps not surprising; we can imagine that there will be dispersion forces between the large halogen atoms and indeed our gas-phase calculations of the bare heterodimer predict a deviation to perpendicular orientation by a few degrees, which may be attractive enough to bend the complex. Another possibility is that the helium environment may be responsible for this structural change, although this is perhaps less likely as the interaction strength with helium is much lower than that between the molecules \cite{calvo_coating_2015,calvo_possible_2016}. Alternatively, an unlikely but possible explanation is that the global minimum energy structure is perfectly perpendicular, however the bent structure forms a local minimum that the complex is trapped in at the 0.37 K temperature of the droplets. We can imagine that as the two molecules approach each other, it is rather infeasible that they do so with a precisely perpendicular angle. If a local minimum does indeed exist, the complex will likely end up trapped there. Frozen local minima of complexes in helium droplets have previously been seen, for example in chains of HCN molecules \cite{nauta_nonequilibrium_1999} and in \ce{Br-HCCCN-Br}~\cite{B611340K}.

\subsection{Coulomb Explosion imaging for structure determination}

How useful is CEI of aligned molecules as a tool for structural determination of loosely bound complexes? Compared to spectroscopic techniques, the structural accuracy for CEI is still much lower. Frequency-resolved spectroscopic tools can provide bond lengths and angles with sub-pm precision, while CEI is restricted to a more qualitative overview of the complex structure. Clearly, if high-resolution spectroscopy is available, then it is the best tool for determining static structures. However, for more complex molecules and complexes, high-resolution spectroscopy is not an option, and often dynamics are more interesting than static structures. It is for these cases where CEI holds the advantage.

Spectroscopic structural determination of complexes is usually accomplished by studying some combination of vibrational and rotational lines, using (for example) ultraviolet, infrared, microwave or rotational coherence spectroscopies~\cite{hobza_world_2006}. However, the applicability of these techniques is often limited. If vibrational transitions are to be used, then the spectrum must be sufficiently uncongested for clear assignment of the spectral lines. This tends to rule out all but simple or highly symmetric complexes. Rotational spectroscopy tends to have clearer assignments for isolated molecules, but inside helium droplets the spectra, and thus the rotational constants, are dominated by interactions with the helium solvent, rendering attempt to extract structural information infeasible~\cite{chatterley_rotational_2020}.

CEI is not the only non-spectroscopic structural tool: recently, the alternative approaches of electron or x-ray diffraction have shown potential for solving helium embedded structures~\cite{he_electron_2016,lei_electron_2020,zhang_electron_2020,a.ikkanda_exploiting_2016,ihee_ultrafast_2005,spence_x-ray_2012,gomez_shapes_2014}. These diffraction experiments (which would also benefit from alignment) have the possibility of directly reporting atomic positions. This can be achieved with free electron lasers~\cite{Neutze2000,Chapman2010,PhysRevLett.112.083002,Ayyer:21}, laser-induced electron diffraction (LIED)~\cite{blaga_imaging_2012} and laser-assisted electron diffraction (LAED)\cite{doi:10.1063/1.4863985}, which all permit a time resolution comparable to that of the current work but with a much higher structural resolution. However, LIED and LAED have so far only been particularly successful on small molecules~\cite{blaga_imaging_2012,pullen_imaging_2015,doi:10.1063/1.4863985,doi:10.1063/1.5093959} and have to our knowledge not yet been applied in helium droplets. The presence of the helium solvent can also be problematic as many events may originate from the helium itself instead of the molecules of interest~\cite{doi:10.1021/acs.jpclett.9b03603} making the scattering diffraction profile of the molecules harder to resolve. The latter will also increase in complexity as larger molecules are being studied which could also limit the technique.

However, neither issue is an inherent limitation for CEI. Molecular complexity certainly hampers CEI, but addition of 'tracer' atoms, such as the Br atoms in this work, allow us to narrow in on only small segments of the molecular structure. Often, a few key parameters are the most interesting for telling the story of a complex structure, and it is in these cases that CEI may be useful. When performed inside a helium droplet, the helium blurs the recoil of fragments as they leave the droplet, but it does not otherwise affect the measurement, unlike it does for rotational spectroscopies~\cite{christensen_deconvoluting_2016}. Hence, CEI may be the correct choice for measuring gross structure of complexes inside helium droplets. Additionally, as demonstrated here, CEI with a 3D detector is able to measure multiple structures simultaneously~\cite{slater_coulomb-explosion_2015}. For helium droplets, this is a particularly powerful feature as the statistical nature of the pickup process naturally means that a mixture of complexes is created when one works under conditions of moderately high signal.

The greatest promise of laser-induced CEI for structure determination undoubtedly lies in femtosecond dynamics. As the duration of the probe laser pulse is tens of femtoseconds, it interacts on a timescale where most molecular motions are frozen.  Intramolecular motions of molecules could be observed, as was previously demonstrated for watching the torsion motion of gas phase biphenyl molecules~\cite{hansen_control_2012}. Even more exciting, one could in principle use femtosecond CEI to watch a bimolecular reaction inside a helium droplet. If one created a complex with a pre-reactive geometry that undergoes bimolecular photoreaction, CEI could be used to study formation of chemical bonds in real-time. In this instance, the lack of spatial resolution may be of little concern if the amplitude of the motion is large enough during the creation of bonds between the molecules. Such a study would represent a huge leap in the ability to monitor chemical reactions.

The primary drawback of CEI is that the fragmentation pattern when the complex undergoes Coulomb explosion radically affects the interpretation. If one would know the fragmentation pathway, the uncertainty we quote could be lowered as we could more reliably know how much angular spread could be expected from the Coulomb explosion. This pathway choice can theoretically be overcome: in the limit of extremely high charge states, cations are reduced entirely into atomic fragments, and hence only one fragmentation pattern is possible \cite{zhou_coulomb_2020}. For large systems, this requires 10s of positive charges in total, which is not easily obtained with a standard femtosecond laser system. However, x-ray free electron lasers can produce intensities high enough to reach these charge states \cite{rudenko_femtosecond_2017}, and can be synchronized with an optical alignment laser, hence XFEL facilities may be where the next generation of these experiments are performed. In principle, these could even be combined with x-ray diffraction experiments, for a very comprehensive view of dimer structures.

\section{Acknowledgments}
We acknowledge support from the following three funding sources: the European Union's Horizon 2020 research and innovation programme under the Marie Sklodowska-Curie Grant Agreement No 674960  "Angular Studies of Photoelectron in Innovative Research Environments" (ASPIRE)  and No 641789 "Molecular Electron Dynamics investigated by Intense Fields and Attosecond Pulses" (MEDEA), a Villum Experiment Grant (No. 23177) and a Villum Investigator grant (No. 25886) from The Villum Foundation.

\appendix
\setcounter{figure}{0}    
\setcounter{table}{0}    
\makeatletter 
\renewcommand{\thefigure}{A\@arabic\c@figure}
\renewcommand{\thetable}{A\arabic{table}}
\makeatother

\section{Mass spectrum}
\label{sec:MassSpectrum}

\begin{figure}[bt]
\centerline{\includegraphics[width = \columnwidth]{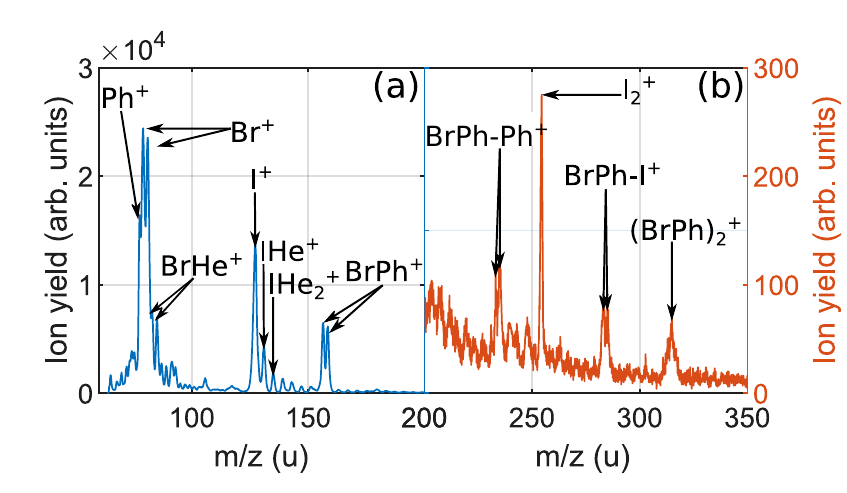}}
\caption{Mass spectrum of He droplets doped with both \ce{I2} and \ce{BrPh} molecules, shown for (a) 60--200 u and (b) 200--350 u. Masses below 60 u were not recorded due to the choice of detector gating.
\label{MASSTOT}}
\end{figure}

The full mass spectrum is shown in \autoref{MASSTOT} with the labeling of relevant ionic fragments. Masses below 60 u were not recorded due to detector gating, and no significant peaks lie above 320 u with the doping conditions used.

\section{Determination of peak centers in covariance maps}
\label{sec:PeakCenter}
\begin{figure}[bt]
\centerline{\includegraphics[width = \columnwidth]{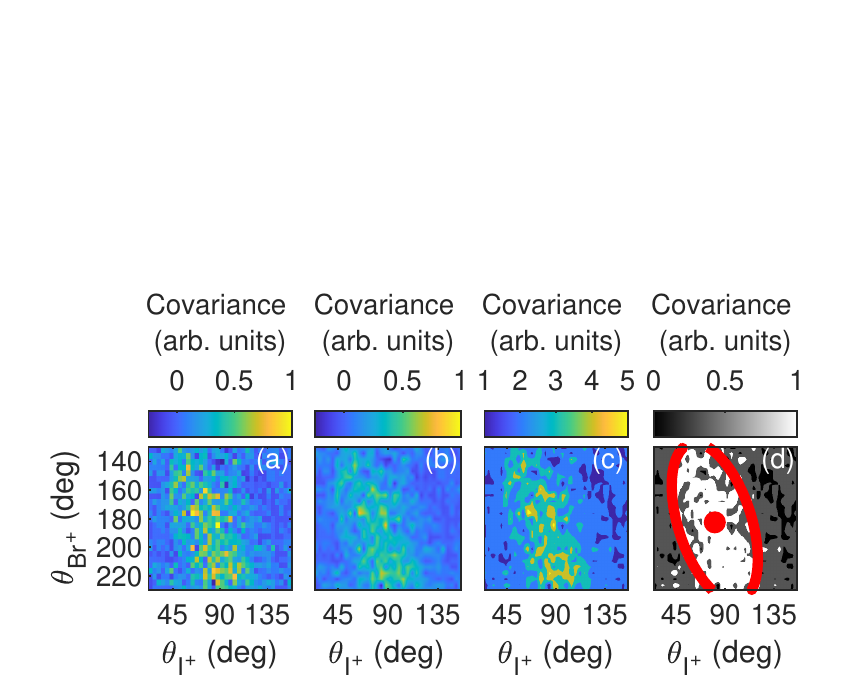}}
\caption{The method used to determine the centers of islands in the $\theta_I$--$\theta_{\rm Br}$ covariance map. The raw image (a) is interpolated (b), binned (c), and then the center is determined by weighted centroid (d). See text for details.
 }
\label{CENTERS}
\end{figure}

The procedure for the center retrieval of the islands in the ($\theta_I,\theta_{Br}$) covariance map, \autoref{I2BRPHCOV}(b$_2$), is highlighted in \autoref{CENTERS}. Panel (a) shows one island from \autoref{I2BRPHCOV}(b$_2$) in the main text. First, the image is interpolated to smooth its shape and facilitate the clustering procedure, using a sampling of 2500 points divided by the initial number of points on each axis. This gives the interpolated image shown in panel (b). Next,
a k-means clustering procedure is applied on the interpolated image to generate multiple clusters according to its intensity. It results into a binning of the image into multiple intensity bins, resulting in the image in panel (c). The binned image is then sent into the Matlab regionprops function. The function will only use the points  weighted by their intensity bins belonging to the clusters associated to the island (here from 3 to 5) to determine the center and bounding ellipse. The output is the red ellipse displayed with a dot at its center, shown in panel (d). Each island is treated separately to facilitate the clustering since the intensity might vary from one island to another.
This methodology has been applied for three different angular binsizes, $2^\circ$, $3^\circ$, $4^\circ$ and for 3 different island selection shown in \autoref{Figure:SelectIsland}. The centers given in \autoref{centreTAB} are the mean positions of each island over all possible configurations.

\begin{figure}[bt]
\centerline{\includegraphics[width = \columnwidth]{cropping.pdf}}
\caption{Different island selections labeled (a), (b), (c) for an angular binsize of $4^\circ$. On the $\theta_{\textrm I^+}$, the thresholds points are ($340^\circ$ - $20^\circ$) and ($160^\circ$ - $200^\circ$) for all panels. On the $\theta_{\textrm {Br}^+}$ axis, the threshold points are, (a): ($70^\circ$ - $110^\circ$) and ($0/360^\circ$), (b): ($70^\circ$ - $110^\circ$) and ($250^\circ$ - $290^\circ$), (c): ($50^\circ$ - $130^\circ$) and ($230^\circ$ - $310^\circ)$. The axis $\theta_{\textrm {Br}^+}$ has been shifted by $-90^\circ$ to facilitate the perception of all four islands.}
\label{Figure:SelectIsland}
\end{figure}

\section{Structures of the \ce{BrPh}--\ce{I2} heterodimer }
\label{sec:BrPh-I2_geom}
The starting geometry for the five potential dimer structures is listed in \autoref{Table:InputGeometry}.
\begin{table}[bt]
\begin{tabular}{|c|c|c|c|}
\hline
Atoms & X & Y & Z \\
\hline
Br &  0.0000 & 0.0000 &   1.7947\\
C  &  0.0000 & 0.0000 &  -0.1006\\
C  &  1.2067 & 0.0000 &  -0.7771\\
C  & -1.2067 & 0.0000 &  -0.7771\\
C  &  1.1982 & 0.0000 &  -2.1627\\
C  & -1.1982 & 0.0000 &  -2.1627\\
C  &  0.0000 & 0.0000 &  -2.8569\\
H  &  2.1375 & 0.0000 &  -0.2299\\
H  & -2.1375 & 0.0000 &  -0.2299\\
H  &  2.1372 & 0.0000 &  -2.6985\\
H  & -2.1372 & 0.0000 &  -2.6985\\
H  &  0.0000 & 0.0000 &  -3.9376\\
I  &  0.0000 & 3.5000 &  -1.4728\\
I  &  0.0000 & 6.2000 &  -1.4728\\
\hline
\end{tabular}
\caption{Starting geometry for the classical simulations in the laboratory frame, where Y is the alignment laser polarization axis and Z the laser propagation axis
\label{Table:InputGeometry}}
\end{table}

For each structure, a charge and a fragment distribution is assumed. The values used in this work are shown in \autoref{Table:Conf}. The charges refer to the electrical charge (in units of $e$) on each atom while the index indicates which atoms belong to the same fragment. Each fragment is considered as a rigid body, such that the distance between its atomic components is fixed throughout the simulation.

The velocity of the iodine ion furthest from the benzene plane and of the bromine ion are determined after 10 ps. A
  distribution of velocity vectors in the laboratory frame is then generated
  taking into account the free rotation of the dimer around the
  alignment laser polarization axis (1D alignment) and the reflection symmetry in a
  plane orthogonal to this axis (no orientation). A Gaussian spread is applied to
  the velocity vectors of both the \ce{Br+} and \ce{I+} ions to simulate the finite degree of alignment of the dimer. The result is a blur of the diagonal line in the ($\theta_{\textrm {Br}^+}$,$\theta_{\textrm I^+}$) covariance map. The effect of the non-axial recoil of the fragment ions is accounted for in a similar manner leading to further blurring of the diagonal line in the covariance map. The observed angular covariance maps shown in \autoref{I2BRPHCOV}(b$_2$), are best reproduced by Gaussian distributions with widths (standard deviations) of 20$^\circ$, 17.5$^\circ$ and 12.5$^\circ$, for the non-perfect alignment and non-axial recoil of the \ce{Br+} and \ce{I+} ions, respectively.
  
The velocities vectors of the \ce{Br+} and \ce{I+} fragments are then projected onto a 2D plane, that either contains the alignment polarization axis or is orthogonal to it in order to represent the two cases shown in \autoref{I2BRPHCOV}. The covariance maps were simulated for a set of $2\times10^6$ events.

\begin{table*}[bt]
\begin{tabular}{|c|c|c|c|c|c|c|c|c|c|c|}
\hline
Atoms &  \multicolumn{2}{c|}{(a)} &  \multicolumn{2}{c|}{(b)} &   \multicolumn{2}{c|}{(c)} &   \multicolumn{2}{c|}{(d)}&   \multicolumn{2}{c|}{(e)}\\
    \cline{2-11}
& Index & Charge & Index & Charge & Index & Charge & Index & Charge & Index & Charge \\
\hline
 Br &  1 &  1         &  1 &  1        &  1 &  1        &  1 &  1        &  1 &  1        \\
 C  &  2 &  0         &  2 &  0        &  2 &  0        &  2 &  0.870700 &  2 &  1.071778       \\
 C  &  3 &  0.190321  &  3 &  0.651576 &  3 &  0.615844 &  2 & -0.664491 &  2 & -0.594518 \\
 C  &  4 &  0.190321  &  4 &  0.651576 &  4 &  0.615844 &  2 & -0.664491 &  2 & -0.594518  \\
 C  &  3 &  0.190321  &  3 &  0.651576 &  5 &  0.615844 &  2 &  0.258182 &  2 &  0.294991 \\
 C  &  4 &  0.190321  &  4 &  0.651576 &  6 &  0.615844 &  2 &  0.258182 &  2 &  0.294991  \\
 C  &  5 &  0.615844  &  5 &  0.615844 &  7 &  0.615844 &  2 & -0.362198 &  2 &  0.022590  \\
 H  &  3 &  0.309679  &  3 &  0.348424 &  3 &  0.384156 &  2 &  0.361166 &  2 &  0.410356  \\
 H  &  4 &  0.309679  &  4 &  0.348424 &  4 &  0.384156 &  2 &  0.361166 &  2 &  0.410356  \\
 H  &  3 &  0.309679  &  3 &  0.348424 &  5 &  0.384156 &  2 &  0.168724 &  2 &  0.218702  \\
 H  &  4 &  0.309679  &  4 &  0.348424 &  6 &  0.384156 &  2 &  0.168724 &  2 &  0.218702 \\
 H  &  5 &  0.384156  &  5 &  0.384156 &  7 &  0.384156 &  2 &  0.244336 &  2 &  0.246571  \\
 I  &  6 &  1         &  6 &  1        &  8 &  1        &  3 &  1        &  3 &  1 \\
 I  &  7 &  1         &  7 &  1        &  9 &  1        &  4 &  1        &  4 &  1 \\
\hline
\end{tabular}
\caption{Charge (in units of $e$) and index parameters used in the simulations
\label{Table:Conf}}
\end{table*}

\section{Polarizability tensors and final structures of the \ce{BrPh}--\ce{I2} heterodimer}
\label{sec:BrPh-I2_pol}

\begin{table}
\centering
\begin{tabular}{cc}
\hline
\hline
Conformation & Polarizability tensors\\							
							\hline
(a) & $\begin{pmatrix} 138.95 \\ 185.38 \\166.99\end{pmatrix}$ \\
(b) & $\begin{pmatrix} 132.45\\179.64\\164.06 \end{pmatrix}$ \\
(c) & $\begin{pmatrix} 132.71 \\ 177.63 \\163.94 \end{pmatrix}$ \\
(d) & $\begin{pmatrix} 132.77\\178.26\\163.79 \end{pmatrix}$ \\
(e) & $\begin{pmatrix} 131.6\\ 188.03\\158.3\\\end{pmatrix}$ \\
\hline
\hline
\end{tabular}
\caption{\label{table:Polarizability} Table listing polarizability tensors for each trial structures presented in \autoref{HETEROSTRUCTURES} . The polarizability components are expressed in $a_0^3$ and are listed in the (x,y,z) order.}
\end{table}

The classical simulations of the angular covariance maps identified the structure of the dimer relative to the aligned axis, i.e. the MPA. Thus, these simulations allow us to determine the angle, $\theta_1$, between the I--I axis and the alignment pulse polarization, as well as the angle, $\theta_2$  between the C--Br axis and the alignment pulse polarization. These angles are listed in \autoref{table:AngleClassicalQuantum}  for each of the five dimer structures in the columns labeled 'classical'. 

For comparison, we calculated the polarizability tensor of the five dimer structures with the $\omega$B97X-D method~\cite{B810189B} with Def2QZVPP~\cite{B508541A} as a basis set. The results, expressed in the frame where the tensor is diagonal, are given in \autoref{table:Polarizability}. These calculations enable a determination of $\theta_1$ and $\theta_2$ and these values are listed in \autoref{table:AngleClassicalQuantum}  in the columns labeled 'quantum'. It can be seen that the classical simulation only agrees with the result from the quantum chemistry calculation for structures (b), (c) and (d).

\begin{table*}[bt]
\begin{tabular}{|c|c|c|c|c|c|c|c|c|c|c|}
\hline
Angle &  \multicolumn{2}{c|}{(a)} &  \multicolumn{2}{c|}{(b)} &   \multicolumn{2}{c|}{(c)} &   \multicolumn{2}{c|}{(d)}&   \multicolumn{2}{c|}{(e)}\\
    \cline{2-11}
& Quantum & Classical & Quantum & Classical &  Quantum & Classical &  Quantum & Classical &  Quantum & Classical \\
\hline
 $\theta_1$  &  14.0\degree &  4.2\degree       &  $\bm{5.1^\circ}$   &  $\bm{3.7^\circ}$        &   $\bm{4.7^\circ}$   &  $\bm{3.3^\circ}$  &  $\bm{6.4^\circ}$   &  $\bm{3.6^\circ}$  &  23\degree &  5.6\degree        \\
 $\theta_2$   &  66.6\degree &  66\degree         &  $\bm{91.9^\circ}$   &  $\bm{83.0^\circ}$    &  $\bm{79.2^\circ}$   &  $\bm{80.6^\circ}$  &  $\bm{77.0^\circ}$   &  $\bm{79.8^\circ}$ &  118.2\degree &  89.6\degree       \\
\hline
\end{tabular}
\caption{Predicted angles between the I--I axis and the C--Br axis with the alignment laser polarization axis for quantum chemical calculations through the polarizability tensor and classical calculations from the recoil of the charged ionic fragments. Bold values refer to conformations that show similar angles between the polarizability calculation and the least square fitting procedure.
\label{table:AngleClassicalQuantum}}
\end{table*}

\section{Structures of the \ce{(BrPh)_2} homodimer}
\label{sec:BrPh_geom}

We simulated Coulomb explosion of the four structures of the homodimer shown in \autoref{BRCONF}. For each structure we calculated the polarizability tensor and determined the recoil velocity of the \ce{Br+} ions (at 10 ps after the probe pulse where the recoil velocity has reached its final direction). The simulations were done using fragmentation pathways (a), (b) and (c) of \ce{BrPh} listed in \autoref{sec:heterodimer} in the main text. Only dimer structure 1 leads to \ce{Br+} recoil directions that are consistent with the experimentally observed back-to-back recoil. 
The other three structures lead to recoil patterns that deviate significantly from the experimental findings when the alignment laser polarization is either parallel or perpendicular to the detector plane.

\begin{figure*}[bt]
\centerline{\includegraphics[width = 2\columnwidth]{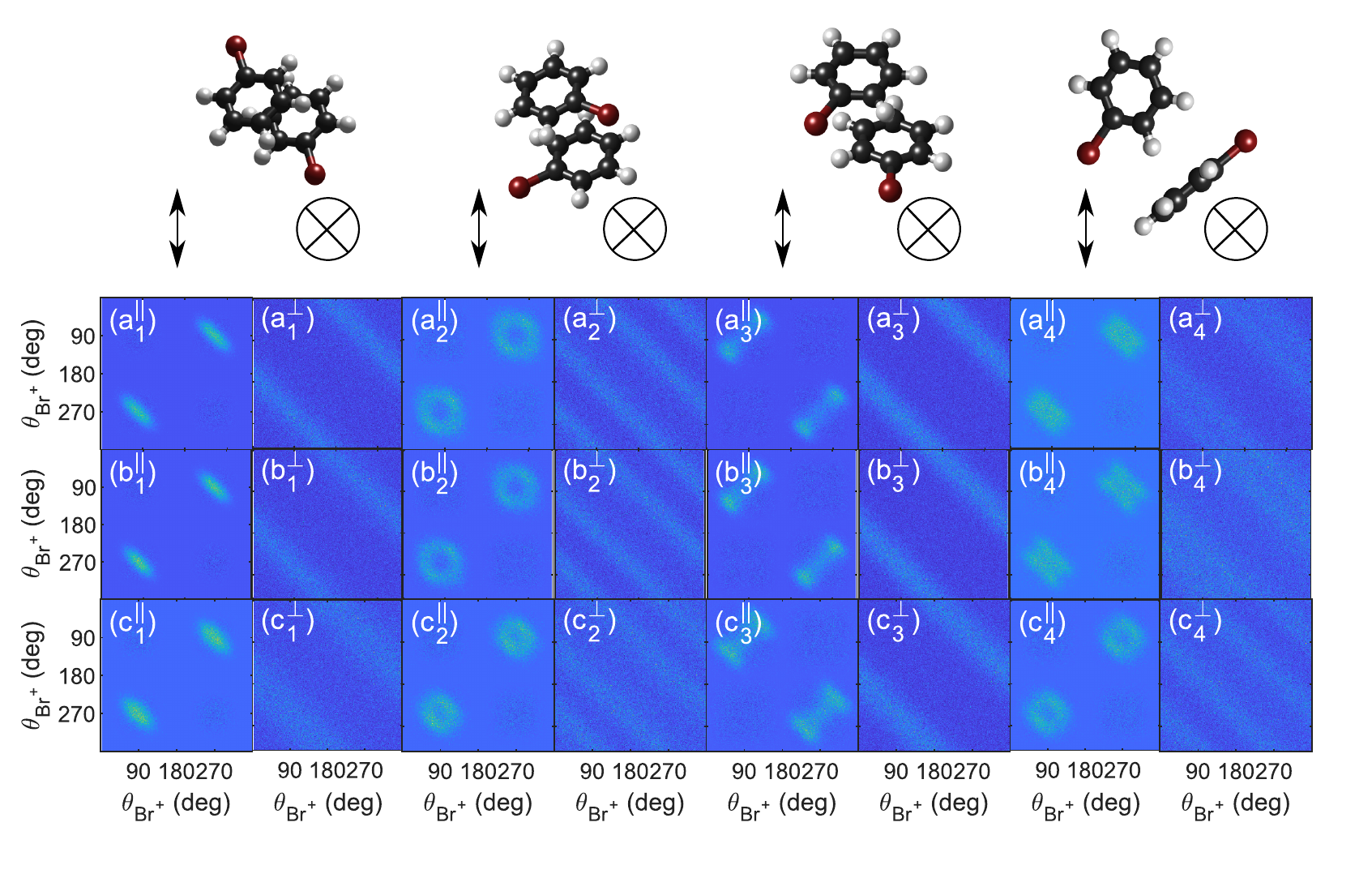}}
\caption{Angular covariance for \ce{Br+} ions for different conformations and charge distributions. The letter labels are associated to the fragmentation of the \ce{C6H5Br} presented in the main text while the numbers refer to each conformation shown in \autoref{BRCONF} and represented on top of the figure. The parallel and the perpendicular sign indicate the relative angle between the polarization of the alignment laser and the detector plane. The standard deviation used for misalignment and non axial recoil is respectively 10$^\circ$ and 10$^\circ$.
\label{BrCov}}
\end{figure*}
\section{Polarizability tensors and structures of the \ce{(BrPh)_2} homodimer}
\label{sec:BrPh_pol}
\begin{table}
\centering
\begin{tabular}{cc}
\hline
\hline
Conformation & Polarizability tensors\\							
							\hline							
1 & $\begin{pmatrix} 223.77 \\ 172.13 \\125.14\end{pmatrix}$ \\
2 & $\begin{pmatrix} 206.77  \\182.57 \\128.18 \end{pmatrix}$ \\
3 & $\begin{pmatrix} 217.62 \\ 178.41 \\124.67 \end{pmatrix}$ \\
4 & $\begin{pmatrix} 224.32 \\ 170.42 \\141.16 \end{pmatrix}$ \\
\hline
\hline
\end{tabular}
\caption{\label{table:Polarizability_BrPh} Table listing the polarizability tensors for each conformation presented in \autoref{BRCONF}. The polarizability components are expressed in $a_0^3$.}
\end{table}
The polarizability tensors of each final structure are shown in \autoref{table:Polarizability_BrPh}. The tensors are
  expressed in the principal axes of polarizability frame.

\section{Predicted structures from geometry optimization}
\label{sec:theory}

Stable structures for the heterodimer and homodimer were also
theoretically investigated, independently from any experimental input,
from a combination of computational methods. A force field exploration
of the energy landscapes was first conducted using the Amber {\it
  ff}99 molecular mechanics framework, and replica-exchange molecular
dynamics. The partial charges on the bromobenzene molecule were taken
from a density-functional theory calculation at the M06-2X level with
aug-cc-pvDZ basis set using the restrained electrostatic potential
fitting procedure. The candidate structures were then reoptimized at
the DFT level, using the same hybrid functional but with the DGDZVP
basis set that is appropriate for the iodine molecule. The DFT
reoptimizations were conducted using the Gaussian09 quantum chemistry
package~\cite{g09}.

\begin{figure}[bt]
\centerline{\includegraphics[width =
    \columnwidth]{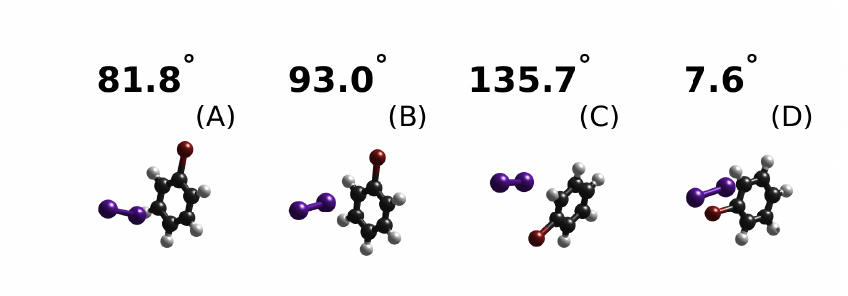}}
\caption{The resultant structures from the geometry optimization of the dimer structure in the gas phase using M06-2x\cite{Zhao2008} with DGDZVP\cite{doi:10.1139/v92-079}. The angle on top of each panel is the relative angle between the I--I axis and the Br--C axis.
\label{Figure:StructureTheory}}
\end{figure}

For the heterodimer, four local-minimum structures were found, as
depicted in \autoref{Figure:StructureTheory}. Their relative
stability can be compared from the binding energy, defined as the
total electronic energy minus twice the monomer energy taken in its
equilibrium geometry. The binding energies thus obtained for the four
heterodimers are (A) 185; (B) 193; (C) 171; and (D) 146 meV,
respectively, confirming that perpendicular configurations for the
iodine molecule are much more stable energetically.

For the homodimer, a similar exploration predicts 3 locally stable
geometries all having the aromatic planes parallel to each other in
shifted and rotated fashions, corresponding to the structures 1, 2,
and 3 in \autoref{BRCONF}. Structure 4 with a T shape is not a local minimum at
the present DFT level. The binding energies obtained for structures 1,
2, and 3 were found to be 332.2, 318.9, and 292.3 meV, respectively.
The prediction that structure 1 is the most stable agrees with the
CEI experiment.
\section*{References}

\bibliographystyle{iopart-num}
\bibliography{Zotero-Dec20}

\end{document}